\documentclass[aps,prl,superscriptaddress,showpacs,floatfix,nofootinbib,notitlepage,twocolumn]{revtex4-1}
\usepackage{amsmath,graphicx,float,hyperref,csquotes,mdframed,appendix,url}

\newcommand{\trento}{T$\mathrel{\protect\raisebox{-2.1pt}{R}}$ENTo}

\begin{document}

\title{Constraining the nucleon size with relativistic nuclear collisions} 

\author{Giuliano Giacalone}
\affiliation{Institut f\"ur Theoretische Physik, Universit\"at Heidelberg,
Philosophenweg 16, 69120 Heidelberg, Germany}

\author{Bj\"orn Schenke}
\affiliation{Physics Department, Brookhaven National Laboratory, Upton, NY 11973, USA}

\author{Chun Shen}
\affiliation{Department of Physics and Astronomy, Wayne State University, Detroit, Michigan 48201, USA}
\affiliation{RIKEN BNL Research Center, Brookhaven National Laboratory, Upton, NY 11973, USA}

\begin{abstract}
The notion of the ``size'' of nucleons and their constituents plays a pivotal role in the current paradigm of the formation and the fluctuations of the quark-gluon plasma produced in high-energy nuclear collision experiments. We report on state-of-the-art hydrodynamic results showing that the correlation between anisotropic flow, $v_n^2$, and the mean transverse momentum of hadrons, $[p_t]$, possesses a unique sensitivity to the nucleon size in off-central heavy-ion collisions.  We argue that existing experimental measurements of this observable support a picture where the relevant length scale characterizing the colliding nucleons is of order 0.5 fm or smaller, and we discuss the broad implications of this finding for future global Bayesian analyses aimed at extracting initial-state and medium properties from nucleus-nucleus collision data, including $v_n^2$-$[p_t]$ correlations. Determinations of the nucleon size in heavy-ion collisions will provide a solid independent constraint on the initial state of small system collisions, and will establish a deep connection between collective flow data in nucleus-nucleus experiments and data on deep inelastic scattering on protons and nuclei.
\end{abstract}

\maketitle

Fluctuations in the energy density distribution of the quark-gluon plasma (QGP) formed in the interaction of two nuclei at relativistic energy originate mainly from the random position of the nucleons that populate the colliding ions at the time of scattering.  State-of-the-art model incarnations of high-energy nuclear experiments require turning the distributions of nucleons that participate in the collision process into continuous density profiles that shape the QGP and serve as initial conditions for its subsequent hydrodynamic expansion. To achieve this, a notion of nucleon \textit{size} is involved, describing the relevant transverse width of nucleons (or their constituents) probed in high-energy experiments. 

Shedding light on the size of nucleons is an important experimental problem as the structure of hadrons, shaped by the non-perturbative dynamics of the strong interaction, remains poorly known. The rest-frame proton \textit{charge radius} has been determined with precision from electromagnetic form factors for both muonic and electronic hydrogen atoms, and is approximately 0.84 fm \cite{Hammer:2019uab}. Inelastic scattering at high energy should instead be sensitive to a smaller \textit{strong radius}. Analysis of diffractive $J/\Psi$ photoproduction at HERA in $e^-p$ collisions suggests that the transverse size of the proton probed in these processes is a gluon radius of order 0.50 fm \cite{Caldwell:2010zza}. A recent study by Kharzeev \cite{Kharzeev:2021qkd} estimates in addition that the proton has a mass radius of about 0.55 fm from data on the threshold photoproduction of vector mesons. 

For about 15 years following the discovery of the important role played by initial-state fluctuations in the phenomenology of heavy-ion collisions \cite{PHOBOS:2006dbo}, hydrodynamic simulations of the QGP have modeled nucleons as boosted two-dimensional Gaussians with a width, $w$, close to 0.4 fm. The corresponding root mean square transverse radius is $ \sqrt{ \langle r^2 \rangle } = \sqrt{2}w \approx 0.56~{\rm fm}$, consistent with the above estimates. These calculations have culminated in the first global Bayesian analysis of Pb+Pb data performed by the Duke group in 2016 \cite{Bernhard:2016tnd}, implementing the \trento{} model \cite{Moreland:2014oya} to initialize the entropy density of the system for the subsequent hydrodynamic evolution, which returns a high-probability [or Maximum A Posteriori (MAP)] nucleon width parameter around 0.45 fm. In addition, a quantitative description of experimental data has been achieved over the past 10 years from hydrodynamic simulations implementing the IP-Glasma model of initial conditions based on the color glass condensate theory of high-energy QCD \cite{Schenke:2020mbo}. This model introduces nucleon sizes consistent with the gluon radius constrained from HERA data  \cite{Caldwell:2010zza,Schenke:2012hg}. The current implementation uses a nucleon width of 0.4 fm, together with sub-nucleonic constituents of width $w_q=0.11$ fm, inferred from incoherent $J/\Psi$ production data \cite{Mantysaari:2016ykx,Mantysaari:2016jaz}.

This consistent state of affairs took a sudden turn in 2019. The second global analysis of the Duke group \cite{Bernhard:2019bmu}, where the \trento{} model is used to compute the initial energy density (as opposed to the entropy density) of the system at $\tau=0^+$, returns in fact MAP nucleon widths of about 0.96 fm, doubling the previous estimate. A similar, more recent analysis by the JETSCAPE collaboration returns even larger values from Pb+Pb data  ($w=0.9$-1.1 fm \cite{JETSCAPE:2020mzn,JETSCAPE:2020shq}). Such nontrivial findings do not have, at present, a clear explanation. Furthermore, analyses of Pb+Pb and also p+Pb data \cite{Moreland:2018gsh,Nijs:2020ors, Nijs:2020roc,Nijs:2021clz} including nucleon constituents with the auxiliary width parameter $w_q$, also return rather diffuse nucleons, $w=0.8$-0.9 fm, albeit with $w_q\approx0.4$ fm. Hydrodynamic results within all these frameworks, e.g., starting with either IP-Glasma initial conditions with sharp nucleons or JETSCAPE MAP \trento{} parameters with very diffuse nucleons return similar final-state observables. There are, thus, stark inconsistencies in the literature concerning the implementation of the nucleon size. 
 
 Anisotropic flow, $v_n$, offers the largest sensitivity to the nucleon width \cite{JETSCAPE:2020shq,Nijs:2021clz}. However, other parameters affect $v_n$ in a similar way, most notably, viscous corrections in the hydrodynamic phase. Apart from improving existing constraints on the QGP initial condition, and permitting us to use heavy-ion collisions as a probe of hadron structure, finding an observable that is strongly sensitive to the nucleon size, and only weakly to medium parameters, would, thus, overcome the degenerate dependence of nucleon width and QGP viscosity on $v_n$, and lead to more precise constraints on transport coefficients and their temperature dependence. In the literature there are, in particular, further inconsistencies concerning the implementation of the specific bulk viscosity of the QGP, whose magnitude (at its peak) currently ranges from very close to zero \cite{Nijs:2021clz} to values that are of the same order as those typically characterizing the specific shear viscosity \cite{Schenke:2020mbo}.  The amount of bulk viscous corrections that models need to implement during the hydrodynamic phase should be connected with the size and the lumpiness of the initial condition, therefore, our understanding of this transport coefficient may largely benefit from an improved knowledge of the nucleon size.
 
 In this Letter, we show that these goals may be readily achieved by means of straightforward experimental measurements, as we point out that the correlation between event-by-event coefficients $v_2$ and $v_3$, and the event-by-event mean transverse momentum of the detected hadrons, $[p_t]$, presents a unique sensitivity to the nucleon width in off-central nucleus-nucleus collisions. We argue, in particular, that existing data on this observable provides nontrivial constraints on the relevant length scales.

We illustrate our point through a simple calculation. We simulate 5.02 TeV Pb+Pb collisions by means of the \trento{} model of initial conditions, where we assume an initial entropy density of the form
\begin{equation}
\label{eq:genmean}
    s(x) \propto  \biggl ( \frac{t_A^p(x) + t_B^p(x)}{2} \biggr)^{1/p} \biggr|_{p=0} = \sqrt{t_A(x)t_B(x)},
\end{equation}
where $x$ is a transverse coordinate and $t_{A(B)}$ is the density of participant matter in nucleus $A(B)$, defined by a superposition of participant densities
\begin{equation}
\label{eq:tA}
    t_A(x) = \sum_j \lambda_j g(x;x_j,w),
\end{equation}
where $g$ is a Gaussian centered around the $j$th participant nucleon. The normalization of each participant, $\lambda_j$, is drawn from a $\Gamma$ distribution of unit mean and variance $1/k$, where $k$ is a real parameter.  In each event we evaluate the energy density of the system from the equation of state of QCD at high-temperature, $e\propto s^{4/3}$. Centrality classes of width 1\% are defined for the simulated events from their initial entropy. For each centrality we calculate the correlation between $v_n^2$ and $[p_t]$ through the Pearson correlation coefficient introduced by Bo\.zek \cite{Bozek:2016yoj}
\begin{equation}
\label{eq:pcc}
    \rho_n \equiv \rho(v_n^2,[p_t]) = \frac{\langle \delta v_n^2 \delta [p_t] \rangle}{\sqrt{\langle (\delta v_n^2 )^2 \rangle \langle (\delta [p_t])^2 \rangle } },
\end{equation}
where $\delta O = O - \langle O \rangle$ for any observable $O$.  To estimate this observable from initial-state quantities, we follow Ref.~\cite{Giacalone:2020dln} and assume that, on an event-by-event basis, $v_n$ is proportional to the corresponding initial-state anisotropy, $\varepsilon_n$, while $[p_t]$ is proportional to the energy of the system. All proportionality factors then cancel in the ratio of Eq.\,(\ref{eq:pcc}). We leave a default parameter $k=1$ (in fact, very close to that found in the JETSCAPE analysis \cite{JETSCAPE:2020mzn}), and choose $w=0.4,~0.8,~1.2$ fm.
\begin{figure}[t]
    \centering
    \includegraphics[width=\linewidth]{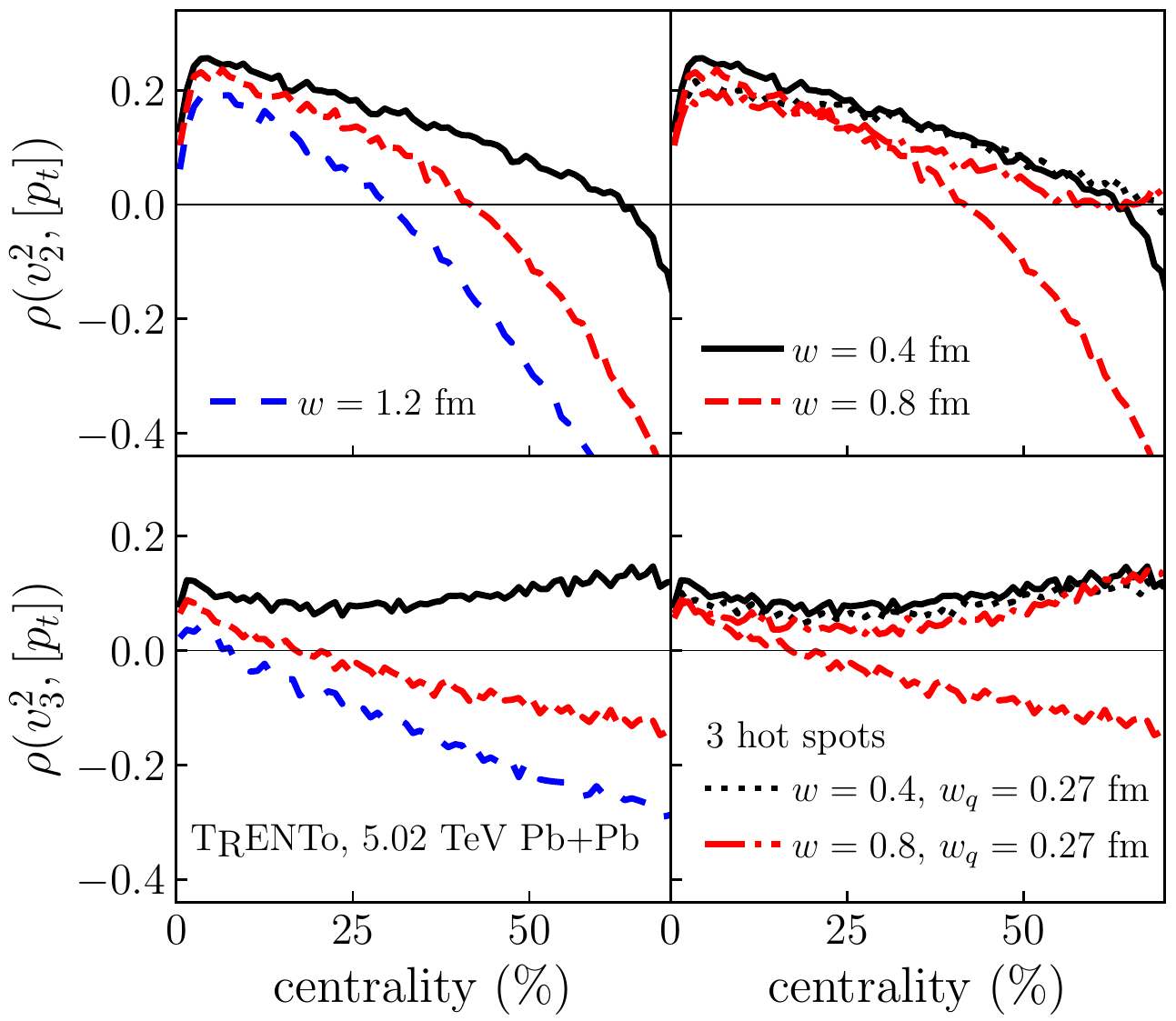}
    \caption{Estimated correlation between $v_n^2$ and $[p_t]$ in the \trento{} model. Left: results for smooth nucleons.  Right: results for smooth nucleons with $w=0.4$ fm and $w=0.8$ fm, and composite nucleons with 3 hot spots with $w_q=0.27$ fm. Different line styles correspond to different parametrizations of nucleon structure. Top: $n=2$. Bottom: $n=3$. }
    \label{fig:1}
\end{figure}

\begin{figure*}[t]
    \centering
    \includegraphics[width=.9\linewidth]{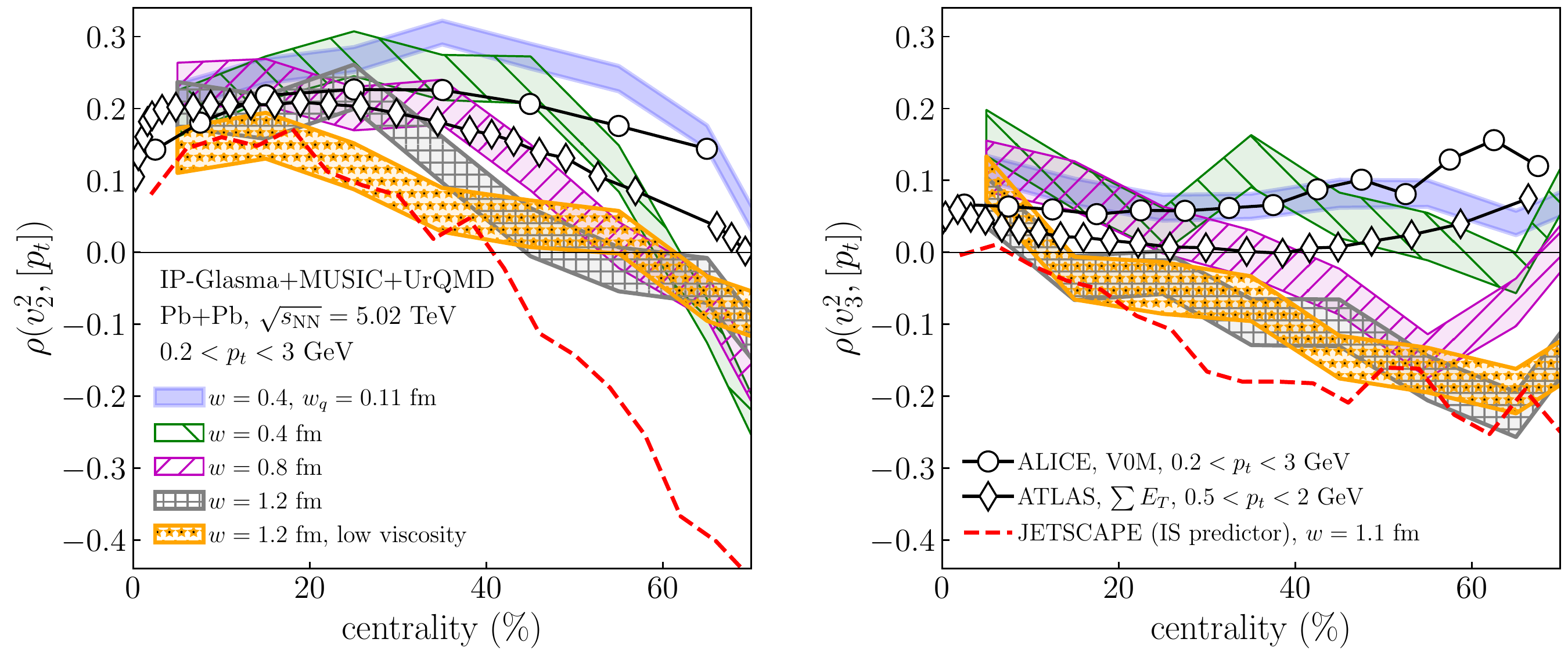}
    \caption{Results of the IP-Glasma+MUSIC+UrQMD framework for $\rho(v_2^2,[p_t])$ (left) and $\rho(v_3^2,[p_t])$ (right). Different types of shaded bands correspond to different values of the nucleon width. The orange band with starry hatches corresponds instead to a calculation with reduced viscosities. The dashed lines are estimators using JETSCAPE initial conditions. Symbols are preliminary ATLAS data \cite{ATLAS:2021kty}  (diamonds) for charged particles with $0.5 < p_t < 2$ GeV, $|\eta|<2.5$ and the centrality defined via $\sum E_T$, and preliminary ALICE data \cite{emil} (circles) for charged particles with $0.2 < p_t < 3$ GeV, $|\eta|<0.8$, and the centrality defined via V0M amplitude.  We note that the larger upper $p_t$ cut implemented in the ALICE analysis explains in part the larger magnitude of their $\rho_n$ compared to ATLAS data for non-central collisions.}
    \label{fig:2}
\end{figure*}

Our \trento{} results are displayed in Fig.~\ref{fig:1}. The panels on the left show dramatic qualitative differences concerning the sign and the centrality dependence of the correlators, depending on the value of $w$. We note that the same observable has been studied at fixed $w$ and fixed $p$ but varying $k$ in Ref.~\cite{Giacalone:2020awm}, as well as at fixed $w$ and fixed $k$ but varying $p$ in Ref.~\cite{Giacalone:2020dln}. In both cases, effects as dramatic as those obtained by varying $w$ in Fig.~\ref{fig:1} are not observed. The panels on the right of the figure present as well results for $w=0.4$ fm and $w=0.8$ fm, but with the inclusion of nucleon constituents (or hot spots). This amounts to replacing the right hand side of Eq.~(\ref{eq:tA}) with $\sum_j \sum_q \lambda_q g(x;x_q,w_q)$, i.e., a sum of Gaussians of width $w_q$ (set to 0.27 fm) centered at $x_q$, which is the coordinate, within the $j$th participant, of the $q$th constituent, sampled from a Gaussian distribution whose width is chosen such that the nucleon is on average a Gaussian of width $w$. We sample 3 hot spots per participant, and we have checked that using 9 does not change our results significantly. Figure~\ref{fig:1} shows, then, that introducing hot spots has a major impact for $w=0.8$ fm, where the scale determining the $\rho_n$ coefficients becomes $w_q$.

The \trento{} estimator captures the centrality dependence and the magnitude of $\rho_n$ obtained at the end of hydrodynamics, up to a correction induced by the upper $p_t$ cut of the particles from which the observable is evaluated, that largely resembles a global rescaling of $\rho_n$ across centralities  \cite{ATLAS:2019pvn,ATLAS:2021kty,emil}. Therefore, we have indeed exhibited an observable with a unique sensitivity to the nucleon (or sub-nucleon) size in nucleus-nucleus collisions.


We demonstrate now that the same conclusion holds in the case of fully-comprehensive simulations of the collision process. We perform simulations of Pb+Pb collisions at $\sqrt{s_{\rm NN}}=5.02$ TeV in a framework that includes $i)$ the scattering of nuclei and the pre-hydrodynamic evolution of the created system within the IP-Glasma model \cite{Schenke:2020mbo}, $ii)$ viscous relativistic hydroydnamic evolution (including shear and bulk viscous corrections, respectively, quantified by $\eta/s$ and $\zeta/s$) performed by the MUSIC hydrodynamic code \cite{Schenke:2010nt,Schenke:2010rr,Paquet:2015lta}, $iii)$ a phase of hadron decays and rescattering that follows the cooling of the QGP, simulated by means of UrQMD \cite{Bass:1998ca,Bleicher:1999xi}. A comprehensive review of our setup can be found in Ref.\,\cite{Schenke:2020mbo}. The simulations of Pb+Pb collisions are repeated for different choices of the nucleon structure. The standard scenario is that of a nucleon composed of three narrow hot spots, each with width $w_q=0.11$ fm \cite{Mantysaari:2016ykx}, whose positions are randomly distributed over the nucleon area following a Gaussian of width $w=0.4$ fm. We have run, in addition, simulations with smooth nucleons with $w=0.4,~0.8,~1.2$ fm. In these different scenarios we have rescaled the energy of the initial profiles (by a factor 1.2) to ensure they yield comparable multiplicities across the centrality range, as entropy production and the inelastic nucleus-nucleus cross section are sensitive to the choice of the nucleon size. Centrality classes are defined from the minimum bias midrapidity charged multiplicity. We calculate our results within 1\% centrality intervals, but since we are limited by statistics, we only show quantities averaged over bins of 10\% width. 

Our results are in Fig.~\ref{fig:2}. The same trends of Fig.~\ref{fig:1} are found, demonstrating that $v_n^2$-$[p_t]$ correlations are driven by the size of the nucleons even in fully-comprehensive calculations. We note that $\rho_2$ for a smooth nucleon with $w=0.4$ fm differs above 50\% centrality from the result with sub-nucleon constituents showing that the details of the sub-nucleon structure are important. The figure reports as well preliminary experimental data from the ATLAS \cite{ATLAS:2021kty} and the ALICE \cite{emil} collaborations at the Large Hadron Collider. Experimental data does not contain evidence of any negative signs for $\rho_n$ across the considered range of centrality. ALICE data is in fair agreement with the IP-Glasma+MUSIC+UrQMD result implementing sub-nucleonic hot spots, computed with the same $p_t$ cuts. Data is in qualitative disagreement with calculations implementing $w=0.8$ fm or larger, and suggests that it is indeed possible to constrain the size of nucleons (or their constituents) from such observations.
\begin{figure*}[t]
    \centering
    \includegraphics[width=.9\linewidth]{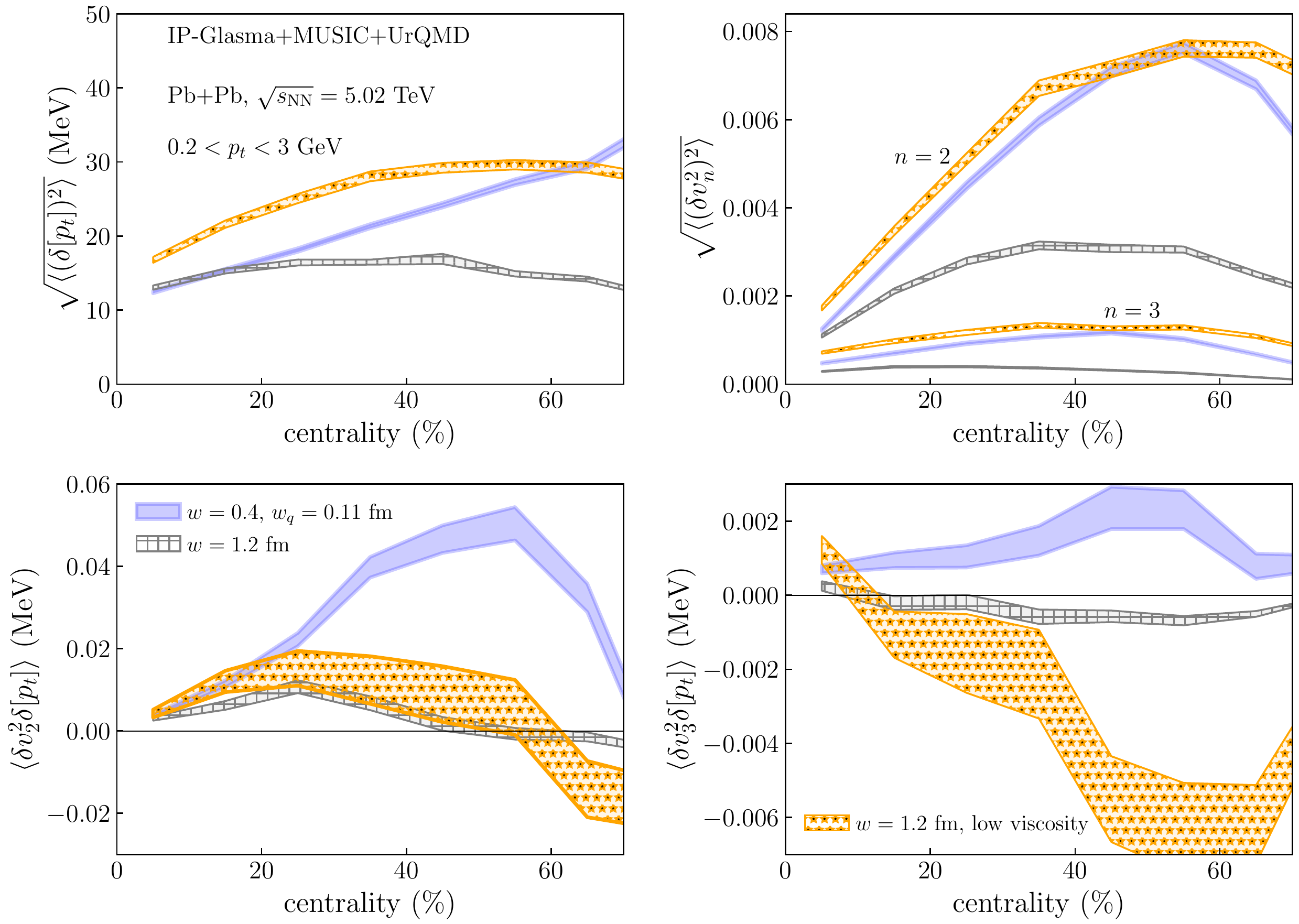}
    \caption{Results from the IP-Glasma+MUSIC+UrQMD framework for the individual components of $\rho(v_n^2,[p_t])$. We show results for two extreme scenarios of nucleon structure,  nucleons with $w=0.4$\,fm with hot spots of size $w_q=0.11$\,fm (blue shaded band), and smooth nucleons with $w=1.2$\,fm (gray hatched band or orange starry-hatched band for the low-viscosity run). Top-left: standard deviation of $[p_t]$ fluctuations. Top-right: standard deviation of $v_n^2$ fluctuations. Bottom-left: covariance of $v_2^2$ and $[p_t]$. Bottom-right: covariance of $v_3^2$ and $[p_t]$.}
    \label{fig:3}
\end{figure*}

To corroborate this point, in the figure we show the result for the same initial-state predictor of Fig.~\ref{fig:1} where the energy and the entropy of the system are computed following the phase of free streaming of JETSCAPE MAP initial conditions with $w=1.1$ fm \cite{JETSCAPE:2020mzn}. We see, as expected, that this calculation yields a result similar to our $w=1.2$ fm curve, although systematically lower in magnitude, possibly due to a residual effect coming from the different scaling of the energy density, namely, $t_At_B$ in IP-Glasma, and $\sqrt{t_At_B}$ in the \trento{} model. This result shows explicitly that, while both the JETSCAPE and the IP-Glasma+MUSIC+UrQMD (in the standard scenario with sub-nucleonic structure) give comparable results for rms flow coefficients, average yields and average momenta, the current JETSCAPE MAP initial condition model leads to a completely wrong description of the measured $\rho_n$ correlators after hydrodynamics. According to our findings it is the large nucleon width inferred by the JETSCAPE analysis that causes this problem. Figure~\ref{fig:2} reports as well on results from a low-viscosity run where we set $\zeta/s=0$ and $\eta/s=0.02$ (and appropriately rescale the initial profiles to account for the loss of viscous entropy production). The result confirms that $\rho_n$ are driven by initial-state fluctuations, and are largely insensitive to medium effects (see also \cite{Magdy:2021ocp}). Their inclusion in future Bayesian analyses will provide a more focused sensitivity to initial-state properties, leading in particular to improved estimates of $w$ (or $w_q$).

To gain a better understanding of the results of Fig.~\ref{fig:2}, we plot in Fig.~\ref{fig:3} the quantities that compose the observable of Eq.~(\ref{eq:pcc}). They are plotted for extreme scenarios of nucleon structure, namely, the standard IP-Glasma model with $w=0.4$ fm and $w_q=0.11$ fm, and the model with the most diffuse nucleons, $w=1.2$ fm. The upper panels of Fig.~\ref{fig:3} display the centrality dependence of the standard deviations appearing in the denominator of Eq.~(\ref{eq:pcc}). A strong depletion of both $v_n$ and $[p_t]$ fluctuations is observed as one sets $w=1.2$ fm. Such dependence is however not as spectacular as that found in the bottom panels of the figure, showing the numerator of Eq.~(\ref{eq:pcc}). Increasing $w$ reduces $\langle \delta v_2^2 \delta [p_t] \rangle$ by orders of magnitude in peripheral collisions, and flips the sign of $\langle \delta v_3^2 \delta [p_t] \rangle$. As argued in Ref.~\cite{Schenke:2020uqq}, the sign of $\rho_2$ should turn from positive to negative at large centrality,  as soon as the almond shape of the system, induced by the collision impact parameter, is lost, and the geometry is dictated by clusters of participant nucleons. Larger nucleon widths reduce the reaction plane eccentricity of the QGP, therefore, it seems plausible that the location of the sign change of $\rho_2$ is observed at lower centralities when $w$ increases. However, we do not have a clear understanding of the great impact of $w$ on $\rho_3$. Nucleon structure properties impact strongly final-state two-particle correlation observables, but mainly in peripheral collisions \cite{Nijs:2021clz}. Our finding that the sign of $\rho_3$ can flip depending on the value of $w$ for centrality percentiles as low as 10\% points, thus, to a new type of sensitivity. Results for $w=1.2$ fm and reduced viscosities are also shown in Fig~\ref{fig:3}. They highlight the remarkable cancellation of viscous effects in $\rho_n$. 

In summary, we have demonstrated that $\rho_n$ coefficients represent powerful probes of the size of nucleons or their constituents in nucleus-nucleus collisions. Existing data on $\rho_n$ supports a picture with a size of order 0.5 fm, in contrast with the findings of recent global Bayesian analyses. In addition, these observables are remarkably insensitive to viscous corrections. Consequently, including $\rho_n$ in future global analyses will lead to improved constraints on both the initial state of the QGP and its transport properties. The possibility of exploiting the great effectiveness of the hydrodynamic framework of high-energy nuclear collisions as a new tool to constrain nucleon structure has two far-reaching implications. First, it provides a means to gauge the initial geometry of small systems, e.g. p+Pb collisions, from the analysis of larger systems alone, which are more robustly understood \cite{Nijs:2021clz}. Secondly, it permits us to assess the consistency of different determinations of the strong nucleon radius relevant to inelastic processes at high energy in different kinds of experiments, most notably, electron-ion collisions to be performed at the future Electron-Ion Collider. 

\bigskip
We thank Wilke van der Schee and You Zhou for useful discussions related to the present topic. G.G. is funded by DFG (German Research Foundation) – Project-ID 273811115 – SFB 1225 ISOQUANT. B.P.S. and C.S. are supported by the U.S. Department of Energy, Office of Science, Office of Nuclear Physics, under DOE Contract Nos.\,DE-SC0012704 and DE-SC0021969, respectively. C.S. acknowledges a DOE Office of Science Early Career Award. This work is in part supported within the framework of the Beam Energy Scan Theory (BEST) Topical Collaboration, and under contract number DE-SC0013460. This research was done using resources provided by the Open Science Grid (OSG) \cite{Pordes:2007zzb, Sfiligoi:2009cct}, which is supported by the National Science Foundation award \#2030508.

\end{document}